\documentclass[aps,11pt,notitlepage]{revtex4}

\usepackage{amsmath}
\usepackage{amssymb}
\usepackage{dcolumn}
\usepackage[dvips]{graphicx}
\usepackage{dcolumn}
\usepackage{bm}
\usepackage[usenames]{color}
\usepackage{verbatim} 

\begin{document}

\title{Controlling spin motion and interactions in a one-dimensional Bose gas}

\author{P. Wicke}
\author{S. Whitlock}
\author{N.~J. van Druten*}

\affiliation{Van der Waals-Zeeman Institute, University of
Amsterdam, Science Park 904, 1098 XH Amsterdam, The Netherlands}
\date{submitted: October 19, 2010}

\begin{abstract}
Experiments on ultracold gases offer unparalleled opportunities to explore quantum many-body physics, with excellent control over key parameters including temperature, density, interactions and even dimensionality. In some systems, atomic interactions can be adjusted by means of magnetic Feshbach resonances, which have played a crucial role in realizing new many-body phenomena. However, suitable Feshbach resonances are not always available, and they offer limited freedom since the magnetic field strength is the only control parameter. Here we show a new way to tune interactions in one-dimensional quantum gases using state-dependent dressed potentials, enabling control over non-equilibrium spin motion in a two-component gas of $^{87}$Rb. The accessible range includes the point of spin-independent interactions where exact quantum many-body solutions are available and the point where spin motion is frozen. This versatility opens a new route to experiments on spin waves, spin-``charge'' separation and the relation between superfluidity and magnetism in low-dimensional quantum gases.
\end{abstract}

\maketitle

Advances in optical and magnetic trapping of ultracold gases have played an essential role in opening up novel avenues in quantum many-body physics by providing experimental access to new physical regimes~\cite{BloDalZwe08}. In particular, one-dimensional (1D) quantum gases, created using optical lattices or atom chips, exhibit a surprisingly rich variety of regimes not present in 2D or 3D~\cite{Ols98,PetSclWal00,RecFedZol03,Khe03,KinWenWei04,ParWidMur04,HofLesSch08,AmeEsWic08}. For example, a 1D Bose gas becomes more {\em strongly} interacting as the density {\em decreases}. Furthermore, the many-body eigenstates and thermodynamic properties of these 1D systems can often be described using exact {\em Bethe Ansatz} methods~\cite{LieLin63,YanYan69,KorBogIze93,Tak99,CauKlaBri09}, and direct comparisons between theory and experiment are possible~\cite{KinWenWei04,KinWenWei05,HofLesSch08,AmeEsWic08,WidTroDem08,LiaRitMue10}. Adding the possibility to dynamically control the strength of atomic interactions, for example via Feshbach resonances~\cite{WidTroDem08,HalGusNag09,HalMarNag10,ChiGriTie10}, there is now a strong impetus to extend these experimental and theoretical studies to non-equilibrium dynamics.

Spinor quantum gases offer the opportunity to study the interplay between internal (spin) and external (motion) degrees of freedom~\cite{SteInoKet98,SchErSen04,FucGanKei05,ZvoCheGia07,KleKolMcC07,WidTroDem08,KroBecSen09,VenGuzSta10,LiaRitMue10}. In this context, strong candidates for experiments are the two magnetically trappable clock states in $^{87}$Rb~\cite{FucGanKei05,KleKolMcC07}, in part because they experience equal trapping potentials and have nearly spin-independent interactions~\cite{HarLewMcG02,MerMerCar07,AndTicHal09}. The drawback is that no convenient Feshbach resonances are available for these states, preventing precise control of the three relevant (inter and intra-state) interaction strengths.

In the two-component (``spin-1/2'') 1D Bose gas, the presence of spin-independent (symmetric) interactions is of particular interest. For all interaction strengths (weak and strong) the dispersion relation of spin waves is quadratic here~\cite{Timmermans98,FucGanKei05}, and the low-energy spin velocity vanishes. As a consequence the usual Luttinger-liquid description~\cite{Giamarchi2004,WidTroDem08,HofLesFis07a,HofLesSch08} cannot be applied. However, it is precisely the point where exact Bethe Ansatz methods can be used~\cite{FucGanKei05,CauKlaBri09}. Furthermore, it is the point where buoyancy effects vanish and in the weakly interacting (mean-field) regime it also lies on the border that separates miscible and immiscible regimes of binary superfluids~\cite{Timmermans98}.

We show that radio-frequency-dressed potentials on atom chips offer a new way to tune the effective interactions in 1D and to control spin motion. We make use of the fact that, for elliptical rf polarizations, different hyperfine states experience different dressed potentials, allowing for state-dependent manipulation~\cite{HofLesFis06}. Here we exploit the dependence of the 1D coupling strength on the transverse confinement frequency $\omega_\perp$~\cite{Ols98}.  State-dependent optical lattice potentials have previously found use for spin-dependent transport and entanglement of atoms~\cite{ManGreBlo03,LeeAndPor07}. More recently, state-dependent microwave dressing was used to generate spin squeezing in 3D Bose-Einstein condensates by varying the wavefunction overlap for two hyperfine states to control collisions~\cite{BohRieTre09,RieBohTre10}, and state-dependent potentials created by combining an optical trap with a magnetic field gradient were used to obtain record low spin temperatures via spin gradient demagnetization cooling of a quantum gas \cite{MedWelKet10}. By tuning the transverse confinement for the two states independently through the rf polarization and amplitude, we show that it is possible to control the interactions in a state- and time-dependent manner. Suddenly changing interactions, combined with the state dependence of the axial trapping then results in dynamical evolution in the spin degree of freedom. In particular, we are able to tune to \emph{(i)} the point where the spin motion is frozen, and \emph{(ii)} the point where the 1D interactions become spin-independent.

\begin{figure}
\includegraphics[width=0.7\columnwidth]{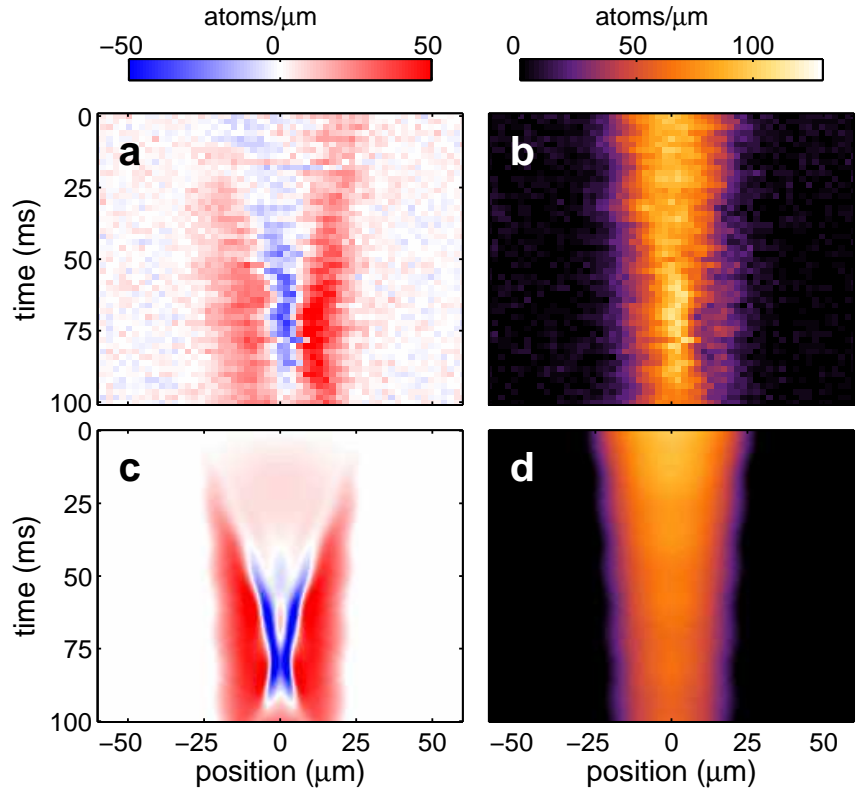}
\caption{[Color] One-dimensional spin dynamics and total density after a sudden transfer of internal-state population, for the case of state-independent trapping potentials. Shown are spin polarization ($n_1-n_2$, left) and total linear density ($n_1+n_2$, right), as a function of axial position and time after the transfer. Top: experiments, bottom: corresponding simulations resulting from integration of two coupled 1D Gross-Pitaevskii equations (GPE). The spin polarization data clearly shows how  $n_2$ is focused towards the center (blue), while $n_1$  moves towards the sides (red); the total density shows little dynamics. Differences between experiment and simulation can be explained by the limited optical resolution of our imaging system and a small tilt of the trap, leading to a slight spatial asymmetry in the experiments.}
\label{fig:data&GP}
\end{figure}

We first discuss our results on the one-dimensional non-equilibrium dynamics for  state-{\em in}dependent potentials, highlighting the importance of small differences in interaction parameters.  The starting point of our experiments is a nearly-pure 1D quasi-condensate in the $|1\rangle\!=|F\!=\!1,m_{f}\!=\!-1\rangle$ state of $^{87}$Rb
in a highly elongated magnetic trap created by an atom chip (see Methods). 
From this initial state, we induce a sudden transition to a coherent superposition of the $|1\rangle$ and $|2\rangle\!=|F\!=\!2,m_{f}\!=\!1\rangle$ hyperfine states via a two-photon pulse, effectively creating a spin-1/2 system \cite{HarLewMcG02,TreHomSte04}.
 The resulting non-equilibrium situation is allowed to evolve for a variable hold time. Subsequently, we directly image the longitudinal distributions, and obtain the linear densities $n_1$ and $n_2$ of the two states along the length of the trap.

 In figure~\ref{fig:data&GP} we present measurements of the evolution of spin polarization ($n_1-n_2$) and the total linear density ($n_1+n_2$) as a function of hold time. The spin pattern shows clear dynamical evolution~[fig.~\ref{fig:data&GP}(a)] whereas the total density remains approximately constant with no significant dynamics [fig.~\ref{fig:data&GP}(b)]. The spin dynamics can be interpreted as a ``focusing'' of state $|2\rangle$ in the presence of state $|1\rangle$, resulting in a negative spin polarization ($n_2>n_1$) toward the center of the trap.

We find good agreement with the experimental data using the coupled 1D Gross-Pitaevskii equations (1D-GPE) with solutions also shown in fig.~\ref{fig:data&GP}(c,d). The 1D-GPE is obtained by integrating the full 3D-GPE over the transverse ground-state wavefunctions~\cite{SalParRea02}, with interaction parameters derived from the intra- and interstate scattering lengths 
taken from ref.~\cite{MerMerCar07}: $a_{11}=100.4\cdot a_0$, $a_{22}=95.00\cdot a_0$ and $a_{12}=97.66\cdot a_0$, where $a_0$ is the Bohr radius. Generalizing for state-dependent harmonic confinement (as will be relevant below) we obtain for the 1D interaction parameters $u_{ij}$:
\begin{eqnarray} u_{11}&=&2\hbar\omega_{\perp,1}a_{11},\nonumber\\
u_{22}&=&2\hbar\omega_{\perp,2}a_{22},\nonumber\\
u_{12}&=&4\hbar\frac{\omega_{\perp,1}\omega_{\perp,2}}{(\omega_{\perp,1}+\omega_{\perp,2})}a_{12},
\label{eq:interactionconstants} \end{eqnarray}
with $\omega_{\perp,j}$ the transverse trap frequency for state $|j\rangle$.
Similarly we use values for the scaled rate constants for inelastic two-body and three-body losses derived from the 3D values in ref.~\cite{MerMerCar07}. The 1D-GPE simulations reproduce the features of the experiment, i.e. absence of dynamics in the total density and the overall structure of the spin dynamics including the time of maximum state separation around $t\approx75~$ms. The decay in atom number on a $\gtrsim 100$~ms timescale is dominated by two-body losses in intrastate interactions and between $|2\rangle$ atoms ($\gamma_{12}$ and $\gamma_{22}$)~\cite{MerMerCar07}.

The rate of spin focusing/defocusing is critically dependent on the precise differences in 1D interaction strengths for the respective internal states, a fact that is readily confirmed by changing these differences in the simulations. The observed general behavior can be understood as follows: in the initial state (an interacting trapped quantum gas in a single internal state in equilibrium) the repulsive  interactions balance the external confining potential. Suddenly transferring a fraction of the population to a second internal state with weaker intra- and interstate interactions results in a net contracting force (a confining effective curvature, $c_2>0$ in equations~\eqref{eq:effectivepotential} below) on the population in this second state that dominates the dynamics in the spin polarization. Because the spin-dependent part of the interactions is relatively small, the dynamics in the {\em total} density are dominated by the (relatively large) average scattering length which remains nearly constant. Hence the total density shows only weak dynamics; the focusing in $n_2$ is accommodated by ``pushing'' $n_1$ to the sides (red in fig.~\ref{fig:data&GP}).

\begin{figure*}[tbp]
\includegraphics[width=0.8\columnwidth]{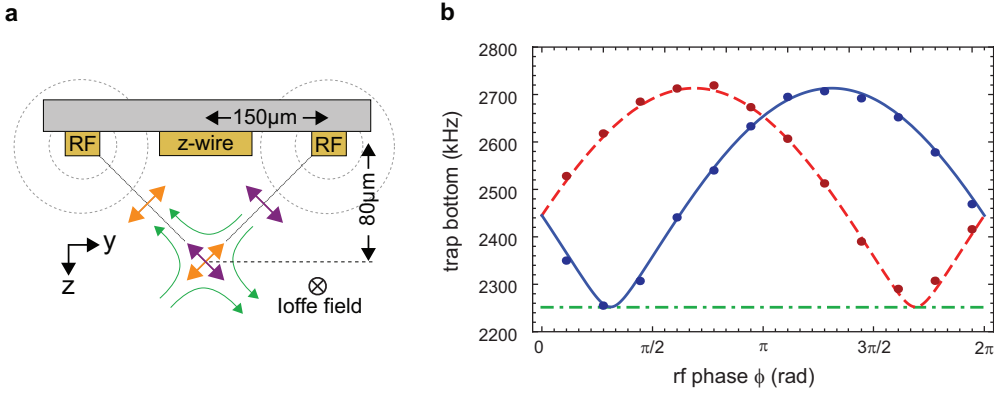}
\caption{[Color] State-dependent potentials. (a) wire geometry used for state-dependent radio-frequency-dressed traps. The static quadrupole magnetic field and the two rf fields are indicated by green, orange and purple arrows respectively. The direction of the bias {\em Ioffe} field which defines the quantization axis is into the plane of the figure. (b) Trap bottom determined via dressed state rf spectroscopy as a function of $\phi$. Data points correspond to the measured trap bottom for state $|1\rangle$ (red) and state $|2\rangle$ (blue). Solid and dashed curves are fits to the data. The dash-dotted green line indicates the fitted trap Larmor frequency $\omega_L/2\pi=2.25$~MHz.}
\label{fig:trapbottom}
\end{figure*}

We now describe the state-dependent radio-frequency-dressed potentials that we use to control the spin motion. We consider near-resonant coupling ($\hbar\omega_{rf}\lesssim g_F\mu_B|B|$) of the rf field with tuneable polarization determined by the relative phase of two independently controlled rf-fields. A cross section of the wire geometry used is shown in fig.~\ref{fig:trapbottom}(a). The fields originate from direct digital synthesis (DDS) supplied currents in two wires neighboring the Z-shaped trapping wire~\cite{EsWicDru10}. With these two fields we can readily control the ellipticity of the total rf field at the trap position by controlling the relative phase $\phi$ of the rf currents in the two wires. This includes linear (horizontal and vertical) and circular ($\sigma^\pm$) polarizations.

The corresponding dressed-state potential for state $|j\rangle$ (with $j=1,2$) has the form $V_j(x,y,z)=((V_0(x,y,z)-\hbar\omega_{rf})^2+\hbar^2\Omega_j^2)^{1/2}$ where $V_0(x,y,z)$ is the bare magnetic (harmonic) potential. The state-dependent part of the potential enters through the coupling Rabi frequency $\Omega_j$~\cite{LesSchSch06,FerGerSpr07}, which acts to weaken the overall confinement near the trap bottom by an amount given by the dressing parameter $\delta_j$. Taking the second derivative of the potential $V_j$ around the origin yields new trap frequencies,
\begin{eqnarray}
\tilde{\omega}_{\perp,\parallel}^2&=\delta_j\omega_{\perp,\parallel}^2,
\textrm{where }
\delta_j&=\Delta/\sqrt{\Omega_j^2+\Delta^2},\label{eq:dressedfreqs}
\end{eqnarray}
with detuning $\Delta=\omega_L-\omega_{rf}$ and Larmor frequency $\omega_L$.

The state-dependent rf potential is characterized using dressed-state rf spectroscopy with a weak additional rf probe~\cite{HofFisLes07,vanEsWhiDru08}. Figure~\ref{fig:trapbottom}(b) shows the measured trap bottom as a function of the dressing phase $\phi$, for $\omega_{rf}=2\pi\times 2.20$~MHz. For $\phi=0.31\pi$ and $\phi=1.68\pi$ the potential is maximally state-dependent, corresponding to the pure circular polarizations $\sigma^-$ and $\sigma^+$ respectively (dressing only state $|1\rangle$ and only state $|2\rangle$, respectively). The potentials are state-independent for linear polarization at $\phi=0,\pi$ (equal dressing of state $|1\rangle$ and $|2\rangle$). The deviation from a simple  $\sin^2(\phi)$ behavior is due to the wire geometry, as the two rf fields are not quite orthogonal at the trap position. A fit to the data (solid and dashed curves) taking into account the wire geometry is used to precisely calibrate all parameters of the rf field coupling (see Methods).

To control the spin motion we turn on the state-dependent dressing, by ramping up the rf currents in $2$~ms with $\omega_{rf}=2\pi\times 2.20$~MHz, directly after preparing the equal superposition of $|1\rangle$ and $|2\rangle$. The ramp time is slow compared to the inverse Larmor frequency and the inverse radial trap frequency, but sudden with respect to any axial motion. We use the two circular rf polarizations  and various rf amplitudes, corresponding to $0.8<\delta_1<1,\delta_2=1$ and $0.8<\delta_2<1,\delta_1=1$. For each time step we extract the widths of the axial distributions in both states.

\begin{figure}[!ht]
\includegraphics[width=0.6\columnwidth,clip=true]{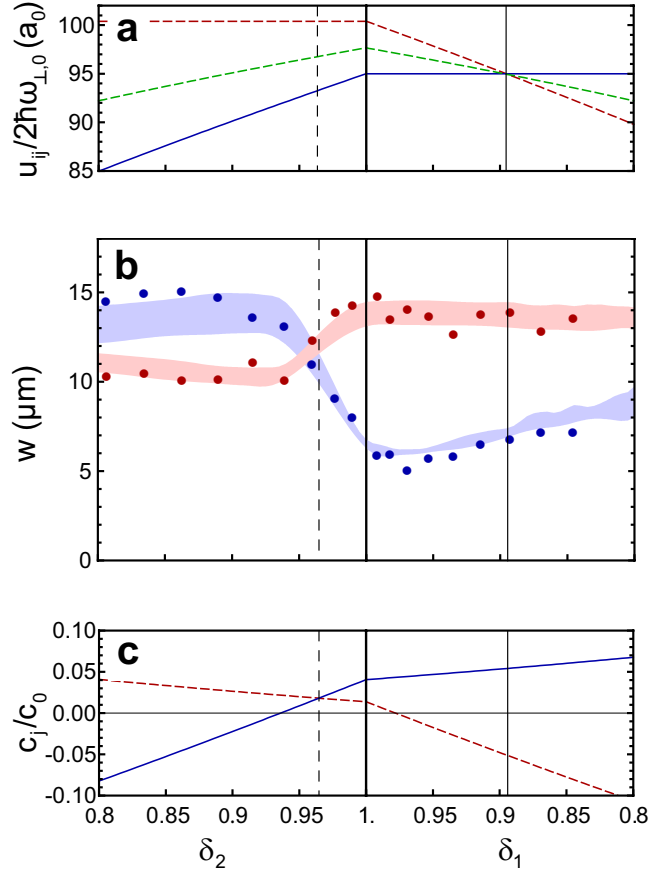}
\caption{[Color] Overview of the possibilities of the state-dependent potentials, as a function of the dressing parameters (left: varying $\delta_2$, with $\delta_1=1$; right: varying $\delta_1$ with $\delta_2=1$).
 (a): 1D interaction strengths, $u_{ij}$ normalised by the bare transverse trap frequency $\omega_{\perp,0}$. (b) Widths of the distribution at $t=44$~ms and (c) scaled effective curvature $c_j/c_0$ at $t=0$. Red indicates state $|1\rangle$ (and $u_{11}$) and blue state $|2\rangle$ (and $u_{22}$) and in (a) $u_{12}$ is indicated in green. The widths in (b) are obtained by a fit to the experimentally measured density profiles (dots) and to GPE simulation (shaded regions). The shaded areas in (b) represent the effect of shot-to-shot atom number fluctuations in the experiment.}
\label{fig:scaledinteractions}
\end{figure}

Results for the full range of dressing parameters are depicted in figure~\ref{fig:scaledinteractions}. Figure \ref{fig:scaledinteractions}(a) shows the calculated interaction strengths taken from equation~\eqref{eq:interactionconstants} as a function of $\delta_1$ and $\delta_2$. We have compared the measured widths of the distributions as a function of time with solutions of the coupled 1D-GPE. These widths and the corresponding simulations for one fixed hold time of 44~ms are shown in figure~\ref{fig:scaledinteractions}(b). The measured widths follow the 1D-GPE simulations closely (taking into account the finite optical resolution), with the biggest uncertainties originating from atom number fluctuations which cause the peak linear density to vary between $70 \mu$m$^{-1}$ and $100~\mu$m$^{-1}$ throughout the entire data set (systematic uncertainty shown by shaded regions). The solid vertical line at $\delta_1=0.895$ indicates the point where the difference in interaction strengths is minimized [fig.~\ref{fig:scaledinteractions}(a)] with $u_{11}$, $u_{22}$ and $u_{12}$ differing by less than 0.05\% (100 times reduction in differences when compared to the unmodified interactions). These conditions are of interest for comparing to Bethe Ansatz solutions which require spin-independent interactions~\cite{FucGanKei05,CauKlaBri09}.

To explain the data we have to consider both the effect of rf-dressing on the collisional interaction strengths as well as the state-dependent modification to the axial potential. A simple analytical description can be obtained using a Thomas-Fermi description near the cloud center where the cloud shape is an inverted parabola. The combination of the state-dependence of the axial trapping frequency and of the interactions can then be expressed as a net harmonic potential characterised by an effective state-dependent curvature $c_j$. We solve for the effective curvatures (Fig.~\ref{fig:scaledinteractions}c) for $t\gtrsim 0$ in our experiments in terms of $\delta_j$, and find
\begin{eqnarray}
\frac{c_1}{c_0}&=&\delta_1-(1-\beta)\sqrt{\delta_1}-\frac{a_{12}}{a_{11}}\frac{2\beta\sqrt{\delta_1\delta_2}}{\sqrt{\delta_1}+\sqrt{\delta_2}}\nonumber\\
\frac{c_2}{c_0}&=&\delta_2-\beta\frac{a_{22}}{a_{11}}\sqrt{\delta_2}-\frac{a_{12}}{a_{11}}\frac{2(1-\beta)\sqrt{\delta_1\delta_2}}{\sqrt{\delta_1}+\sqrt{\delta_2}}
\label{eq:effectivepotential}
\end{eqnarray}
Here the first term on the right-hand side reflects the modification to the external axial potential and the second and third terms deal with the modified interactions $u_{ij}$. The axial curvature of the bare potential is $c_0=m\omega_\parallel^2/2$ and $\beta$ corresponds to the fraction of the population transferred to state $|2\rangle$ ($\beta\approx1/2$ for our experiments).

The dashed line at $\delta_2=0.96$ in fig.~\ref{fig:scaledinteractions} indicates the point where the difference in interactions is compensated by the state-dependent longitudinal potential and $c_1=c_2$. This point is characterized by small equal curvatures of the effective potentials (including interaction energy) for both states [fig.~\ref{fig:scaledinteractions}(c)], which result in frozen spin dynamics. These conditions are important for applications with on-chip atomic clocks, to minimize inhomogeneous broadening due to mean field shifts.  For $\delta_2<0.96$ the difference in interaction strengths is further enhanced and the time evolution of the spin dynamics becomes inverted, with focusing of state $|1\rangle$ while state $|2\rangle$ is pushed outward, as is visible in fig.~\ref{fig:scaledinteractions}(b).

\begin{figure}[!ht]
\includegraphics[width=0.7\columnwidth]{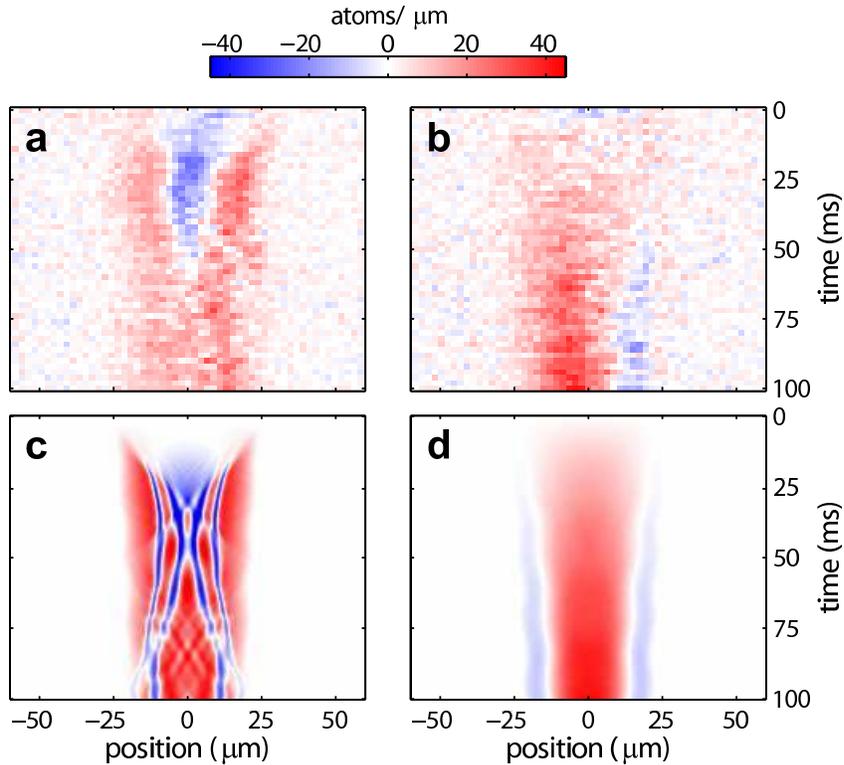}
\caption{[Color]  Spatiotemporal behavior of the spin polarization ($n_1-n_2$) following the sudden transfer to the state-dependent potentials. States $|1\rangle$ and $|2\rangle$ are indicated red and blue, respectively. (a) shows the evolution with rf parameters corresponding to equal inter-atomic interactions ($\delta_1=0.895,\delta_2=1$) and (b) equal effective potentials ($\delta_1=1,\delta_2=0.96$). (c) and (d) show the results of 1D-GPE simulations corresponding to (a) and (b), respectively.}
\label{fig:dressedevo}
\end{figure}

\begin{figure}[!ht]
\includegraphics[width=0.5\columnwidth]{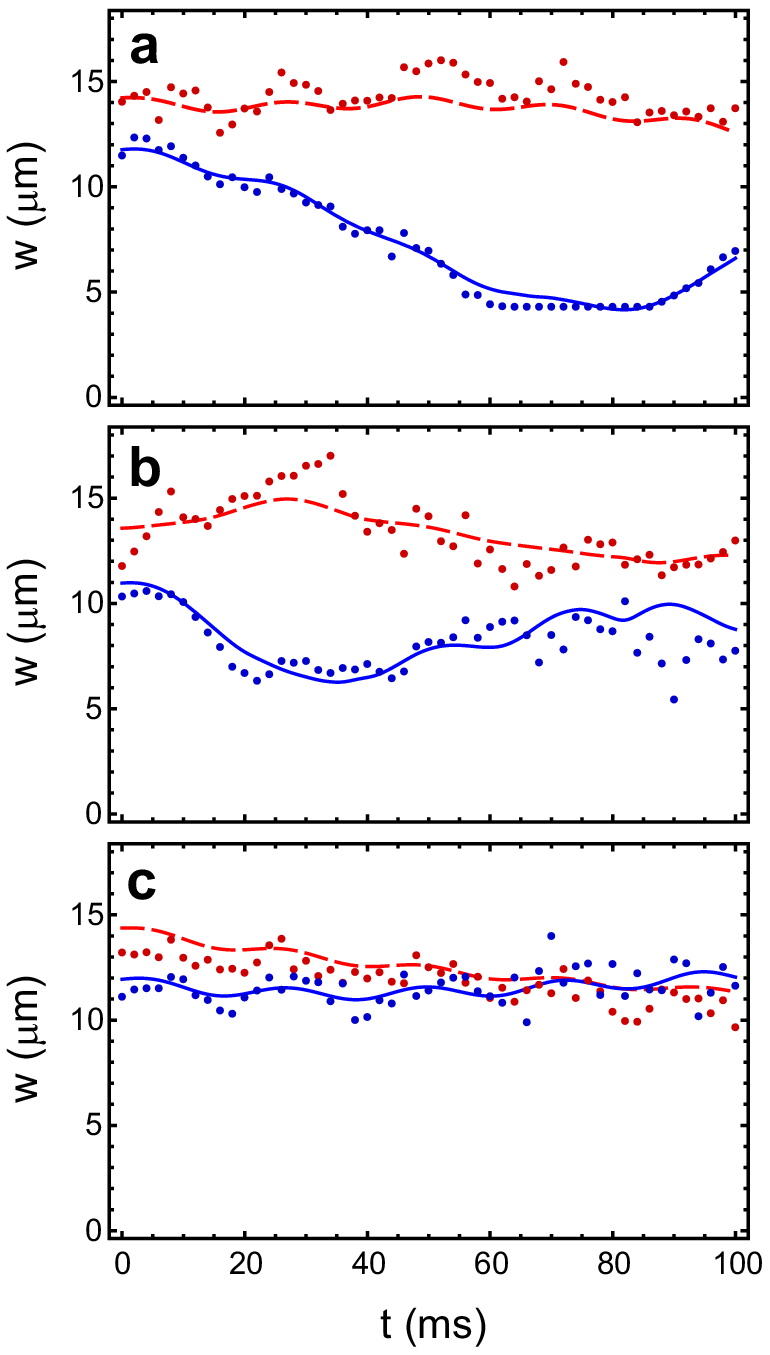}
\caption{[Color]  Widths of the atomic distributions as function of hold time, for the
parameters indicated by the three vertical lines in figure~\ref{fig:scaledinteractions}.
Dots are fits to experimental data, lines are results of 1D GPE solutions. States $|1\rangle$ and $|2\rangle$ are indicated red and blue, respectively. In (a) no rf dressing is applied $\delta_{1,2}=1$, as in figure~\ref{fig:data&GP}. (b) and (c)
correspond to the data in figure~\ref{fig:dressedevo}.
 (b) shows the evolution with rf parameters corresponding to equal inter-atomic interactions ($\delta_1=0.895,\delta_2=1$),
 and (c) equal effective potentials ($\delta_1=1,\delta_2=0.96$).}
\label{fig:widths}
\end{figure}

Figure~\ref{fig:dressedevo} shows the full time evolution of the spin polarization for two selected rf-dressing parameters. The selected cases are: dressing of state $|1\rangle$ alone ($\phi=0.31\pi$, $\delta_1\!=\!0.895$) [fig.~\ref{fig:dressedevo}(a)] and dressing state $|2\rangle$ alone ($\phi=1.68\pi$, $\delta_2\!=\!0.96$) [fig.~\ref{fig:dressedevo}(b)], corresponding to the intersection points in figure~\ref{fig:scaledinteractions}(a) and (c), respectively. Qualitatively state $|2\rangle$ focuses faster with rf dressing applied to state $|1\rangle$, when compared to the case of state independent potentials [fig.~\ref{fig:data&GP}(a)]. Generally, the simulated density profiles reveal a rich and dynamic nonlinear evolution of the spin polarization, reminiscent of filament propagation in  optical systems with competing nonlinearities \cite{CouMys07}. This is clearly visible in figure~\ref{fig:dressedevo}(c) for example.
This detailed structure depends sensitively on the precise values of the dressing. The development and propagation of this fine structure in the spin polarization is partially observed in the experimental data, but is not fully resolved due to the finite imaging resolution. Convolving the simulated profiles with the point-spread function of our imaging system yields excellent agreement with all of the data. With weak dressing of state $|2\rangle$ ($\delta_2\!=\!0.96$) it is possible to freeze spin dynamics altogether such that the two states maintain their overlap and the widths remain constant (apart from a small in-phase quadrupole oscillation and decay from state $|2\rangle$), see Figure~\ref{fig:dressedevo}(b,d). A more quantitative representation of the data,  showing excellent
agreement between experiment and simulation, is given in figure~\ref{fig:widths}, where the widths of the two states are shown for different evolution times. Clearly, the focus point can be identified in figure~\ref{fig:widths}(a) around $t=75$~ms and
in figure~\ref{fig:widths}(b) around $t=30$~ms, whereas no focussing is present in figure~\ref{fig:widths}(c).

We have shown that by introducing a small state-dependence to the radial trapping potential using rf dressing we can precisely tune the 1D interaction parameters in a two-component quantum gas by more than 10\%, over an experimentally significant range. In our experiments this modification competes with the state dependence of the axial trapping and provides a new ``knob'' to control spin motion, leading to tuneable nonlinear behavior.

Our method can be naturally extended in several ways. For instance, control over the interactions {\em without} the accompanying state-dependence of the axial trapping can be obtained by using one-dimensional box-shaped potentials~\cite{EsWicDru10}.  By introducing an additional displacement of the transverse potential in a state-dependent way it is possible to further reduce $u_{12}$, allowing all three interaction parameters to be tuned independently, something that is not generally possible with a magnetically controlled Feshbach resonance.

The observed spin dynamics depend critically on the precise differences in interaction strengths. For $^{87}$Rb, the three relevant scattering lengths are nearly equal and therefore weak dressing is sufficient to tune the system parameters to the point of symmetric interactions or to where the spin dynamics become frozen. Since the rf parameters can be precisely known, such experiments could also allow precision determination of the scattering length differences.
More generally, tuning the system parameters around the point of spin-independent interactions strongly affects the dispersion relation of the spin excitations \cite{Timmermans98}. In particular this allows the spin velocity to be tuned around zero, providing a new handle for the study of spin waves in one-dimensional atomic gases.

Tunable interactions in two-component quantum gases have important applications in the areas of spin-squeezing and quantum metrology~\cite{RieBohTre10,GroZibObe10}, and the ability to control spin motion opens new avenues for future studies of quantum coherence in interacting quantum systems~\cite{WidTroDem08,TreHomSte04,MerMerCar07,AndTicHal09,DeuRamRos10}. Our current experiments are performed in the weakly interacting 1D regime and at low temperature, and we find that a description based on two coupled 1D Gross-Pitaevskii equations is sufficient to describe our data. The methods presented here to tune interactions are not limited to this regime, however. In particular, we plan to apply these methods to systems with stronger interactions (e.g., by lowering the 1D density) and with higher temperatures. This will provide experimental tests of (and challenges to) more sophisticated theoretical methods, for both equilibrium and non-equilibrium phenomena. For instance, it will be possible to experimentally explore predictions of the thermodynamic Bethe Ansatz for the two-component Bose gas~\cite{CauKlaBri09} and to explore quantum quenches in strongly interacting 1D systems by dynamical control over the spin-dependent interactions. Finally, we expect that the experimental control over spin motion and interactions, as demonstrated here, will benefit the
realization of spin-``charge'' separation in a Bose gas \cite{FucGanKei05,KleKolMcC07}.

\begin{acknowledgments}
We thank R. J. C. Spreeuw and J. T. M. Walraven for valuable discussions. We are grateful to FOM and NWO for financial support. SW acknowledges support from a Marie-Curie fellowship (PIIF-GA-2008-220794).
\end{acknowledgments}

\section*{Methods}

\subsection*{Initial-state preparation and coherent spin mixing}

$^{87}$Rb atoms in state $|1\rangle\!=|F\!=\!1,m_{f}\!=\!-1\rangle$
are evaporatively cooled to quantum degeneracy in a highly elongated Ioffe-Pritchard microtrap with trap frequencies of $\omega_\perp/2\pi=1.9$~kHz and $\omega_\parallel/2\pi=26$~Hz. The peak linear atomic density is $n_1\lesssim100~\mu$m$^{-1}$. In this system, both the temperature and chemical potential are small compared to the radial excitation energy ($\mu,k_BT < \hbar\omega_\perp$) and the dynamics are restricted to the axial dimension (1D regime). A coherent superposition of the $|1\rangle$ and $|2\rangle\!=|F\!=\!2,m_{f}\!=\!1\rangle$ hyperfine states is prepared using a resonant two-photon rf and microwave (mw) coupling~\cite{HarLewMcG02,TreHomSte04}. The microwave frequency is introduced via an external antenna while the rf-field is applied directly to the atom chip wires. The measured two-photon Rabi frequency is $\Omega_{c}/2\pi=1.14$~kHz, corresponding to a $\pi/2$-pulse duration of $0.22$~ms. This is fast compared to the timescale for axial dynamics, but sufficiently slow to prevent radial excitations. Coherence times in excess of 1~second have been measured in this setup via Ramsey spectroscopy of dilute thermal clouds.

\subsection*{State-dependent imaging}

The time evolution of the spin distribution is measured by varying the hold time after the $\pi/2$-pulse and performing sequential state-dependent absorption imaging. The atom cloud is released from the trap for 1.0~ms time-of-flight to improve the detection efficiency while preserving the longitudinal distribution. Atoms in state $|2\rangle$ are imaged directly using absorption on the $F=2,F'=3$ transition with an exposure time of $30~\mu$s and an optical resolution of $4.2~\mu$m. State $|1\rangle$ is measured by first removing $|2\rangle$ atoms with a resonant light pulse followed by a 1~ms repumping pulse from $F=1$ to $F=2$. The remaining atoms are then imaged in the same way as for state $|2\rangle$. Due to the extra repumping step we find a 20\% lower detection efficiency for state $|1\rangle$ and a poorer resolution of $\sim 8~\mu$m due to photon recoil, visible in fig.~\ref{fig:widths}. The resulting absorption images are integrated along the radial direction to obtain the linear densities $n_1$ and $n_2$ of the two states.

\subsection*{Characterizing state-dependent potentials}

 The potential energy at the trap bottom is characterized by the onset of loss as a function of probe frequency which we fit to extract $V_j(0,0,0)$.  The measured trap bottom varies with rf phase between 2250~kHz and 2700~kHz corresponding to a maximum Rabi frequency of $\Omega_j/2\pi=450~$kHz.  A fit to the data (solid lines in
 figure~\ref{fig:trapbottom}(b)) taking into account the wire geometry results in
 an accurate calibration of the key experimental parameters, in particular the Larmor frequency  $\omega_L\!=\!2.25~$MHz, rf field amplitudes $b_1=b_2=0.53~$G from the two rf wires and the trap-surface distance of $80~\mu$m.

\end{document}